\documentclass[galaxies,article,accept,oneauthor,10pt,a4paper]{mdpi}
\firstpage{1}
\issuenum{0}
\articlenumber{00}
\pubvolume{6}
\pubyear{2018}
\copyrightyear{2018}
\history{Received: 31 January 2018; Accepted: 12 April 2018; Published: date}
\usepackage{amsmath,amsfonts,amssymb}
\usepackage{graphicx}
\usepackage{aas_macros}
\usepackage{amsfonts}
\usepackage{multirow}

\usepackage{booktabs}
\usepackage{multirow}
\usepackage{soul}
\usepackage{microtype}
\usepackage{upgreek}
\usepackage{textcomp}

\Title{Instrumentation and Future Missions in the Upcoming Era of X-Ray Polarimetry}

\Author{Sergio Fabiani $^{\dagger}$\orcidA{}}

\address[1]{%
INAF-IAPS; sergio.fabiani@iaps.inaf.it; Tel.: +39-06-4993-4450}

\firstnote{\hangafter=1 \hangindent=1.05em \hspace{-0.82em} Current address: via del Fosso del Cavaliere 100, 00133 Rome, Italy}

\abstract{The maturity of current detectors based on technologies that range from solid state to gases renewed the interest for X-ray polarimetry, raising the enthusiasm of a wide scientific community to improve the performance of polarimeters as well as to produce more detailed theoretical predictions.
We will introduce the basic concepts about measuring the polarization of photons, especially in the X-rays, and we will review the current state of the art of polarimeters in a wide energy range from soft~to hard X-rays, from solar flares to distant astrophysical sources.
We will introduce relevant examples of polarimeters developed from the recent past up to the panorama of upcoming space missions to show how the recent development of the technology is allowing reopening the observational window of X-ray polarimetry.}


\begin{document}

\section{Introduction}

X-ray polarimetry still remains nowadays a vastly unexplored field in astronomy, but the recent development of new technologies will allow us in a near future to deeply observe many astrophysical sources with measurements with high significance level. Despite imaging, spectroscopy and timing being well-developed techniques also in X-ray astronomy, polarimetry remains an observational tool that is less advanced with respect to the other wavelengths.
However, many emission processes and physical interactions involve the production of polarized photons also in the X-rays.

In this paper, the science goals of X-ray polarimetry are briefly introduced in Section~\ref{sec:goals}. The basic concepts of polarimetry at high energy and the typical relevant detector systematics are discussed in Sections~\ref{sec:basics} and \ref{sec:systematics}, respectively.
In Section~\ref{sec:techniques}, the polarimetry techniques employed in X-ray astronomy~are discussed in detail as well as the related instrumentation and some relevant polarimetric missions that have recently launched, are scheduled for launch or in assessment phase.

\section{Science Goals of X-Ray Polarimetry}\label{sec:goals}

Many physical processes are responsible for the polarization of X-rays; therefore, polarimetry is a powerful tool to investigate the physics and the geometry of different families of astrophysical sources. Polarimetry adds the degree and the angle of polarization to the space of observable parameters allowing one to add information useful to remove degeneracies in source models and to distinguish among different types of these.

The processes of particle acceleration are among the most relevant phenomena responsible for the emission of polarized X-rays.
The magnetic field plays a crucial role in the particle acceleration, depending on its turbulence level and reconnection events. The knowledge of the magnetic field's geometry and strength is useful to place constraints on the parameters of the models of acceleration mechanism. Synchrotron radiation \citep{Westfold1959} is the emission process fed by particle acceleration in astrophysics sources like Pulsar Wind Nebulae (PWNe) that originate from the interaction of the pulsar magnetic field and the relativistic wind with the surrounding Supernova Remnant (SNR) \citep{Kennel1984a, Kennel1984b}.
They are efficient particle accelerators \citep{Nakamura2007,Volpi2009} and emit synchrotron radiation in the X-ray band. The Crab Nebula is the only astronomical source with a high confidence level X-ray polarimetric measurement~of $19.2\pm1.0\%$ \citep{Weisskopf1978}.
Another class of particle accelerators are Young Supernova Remnants (SNRs) \citep{Helder2012,Bykov2009,Reynolds1981} in which accelerated electrons radiate via synchrotron radiation along shock fronts, producing emitting filaments.
Particle acceleration is responsible also for feeding in the physics of jets (both in galactic and extragalactic sources).
Among Active Galactic Nuclei (AGN), blazars are powered by an efficient accretion onto a supermassive Black Hole (BH) of highly energized and magnetized plasma \citep{Blandford1982}. The~lower~energy peak in their Spectral Energy Distribution (SED) is due to synchrotron radiation and can reside in the X-ray, the optical or the IR energy bands.
The physics of jets and their relationship with the disc and the corona comprise a key topic for polarimetry when studying galactic sources like micro-quasars \citep{Fender2001,Mirabel1999}. Scenarios with
Comptonization of thermal/quasi-thermal disc photons in a hot electron-positron corona compete with synchrotron models of a relativistic jet.

The ejection of a highly-collimated high-speed jets of plasma should be the cause responsible~for the emission of Gamma Ray Bursts (GRB). The mechanisms of the production of the prompt radiation, the energetics of explosions and the role of magnetic fields remain largely unknown.
Particle~acceleration is also responsible for the emission of solar flares originating from magnetic reconnection. Accelerated electrons emits via non-thermal bremsstrahlung \citep{Gluckstern1953} and their emission can~be highly polarized \citep{Zharkova2011}.

Another field of investigation for X-ray polarimetry is comprised of sources, the emission of which originates in strong magnetic fields that channel matter accretion along field lines, creating aspherical X-ray emission and scattering geometries.
X-ray polarimetry is able to probe the origin and structure of the emission from magnetized White Dwarfs (WD) \citep{Rosen1988,Lamb1979} and Neutron Stars (NS) in binary systems. In accreting X-ray pulsars, plasma coupling with a strong magnetic field ($10^{12-13}$~G) is responsible~for the high polarization of X-rays. The emission from compact sources can be polarized due to the different opacity of the plasma to the different states of the polarization \citep{Meszaros1988,Gnedin1974,Ventura1979}.
Polarimetry~in X-rays is also a valuable tool to study vacuum polarization and birefringence through extreme magnetic fields~\citep{Gnedin1978,Ventura1979b,Meszaros1979}.
In accretion-powered millisecond pulsars (AMs) \citep{Wijnands2005,Patruno2012,Poutanen2008}, the hard power-law component in the energy spectra is likely due to Comptonization in a radiative shock surface of the thermal emission from~a hot spot on the neutron star surface. The scattered radiation should be linearly polarized with the polarization degree and angle varying with the phase \citep{Viironen2004}.
X-ray polarimetry offers also the possibility~to study rotation-powered pulsars \citep{Slowikowska2009}, the X-ray emission of which is expected to be highly polarized, as well as~to study ultra-magnetized neutron stars, such as magnetars \citep{Taverna2014,Thompson2002,Turolla2015}.

Scattering in aspherical geometries is responsible for polarizing radiation. This is the case of accreting plasmas in disks, blobs and columns \citep{Sunyaev1985,Meszaros1988,Sazonov2001}. This process is responsible, for example for the polarization of the emission in binary systems (XRBs) and AGNs \citep{Rees1975}.
Comptonization arises from accretion-disk-fed binary sources in which disk photons are Compton scattered by a hot corona~\citep{Haardt1993,Poutanen1993}. Polarimetry can put constraints on the corona and on the inclination angle of the disk \citep{Schnittman2010}. Moreover, also the fraction of the primary emission reflected by the accretion disk itself (Compton Reflection component (CR)) is expected to be highly polarized depending mainly on the inclination angle of the disk \citep{Matt1989}.
The innermost regions of radio-quiet AGN can be seen as scaled-up versions of galactic black hole systems with the hard Comptonization component produced from the thermal UV/soft-X-ray disk component.
Moreover, in addition to the accretion disk, other reflecting regions are present, such~as the dusty torus.
Reflection is also the process responsible for the emission from molecular clouds as for example Sgr B2, which is probably reflecting the past
emission from the central black hole source of our galaxy \cite{Ponti2013,Churazov2002,Koyama1996,Sunyaev1993}, which should therefore have undergone a phase of strong activity about three hundred years ago.

Different fields of fundamental physics can be investigated by means of X-ray polarimetry. Light~bending around BHs is a general relativity effect responsible for the rotation of the polarization vector, which is particularly relevant in strong fields \citep{Stark1977,Connors1980,Dovciak2008,Li2009,Schnittman2009}. Furthermore, new physics, such as Quantum Gravity (QG) \citep{Gambini1999} and the search for axion-like particles \citep{Bassan2010} would be possible.

\section{Polarimetry Basics}\label{sec:basics}

\begin{figure}[H]
\centering
\includegraphics[width=5 cm]{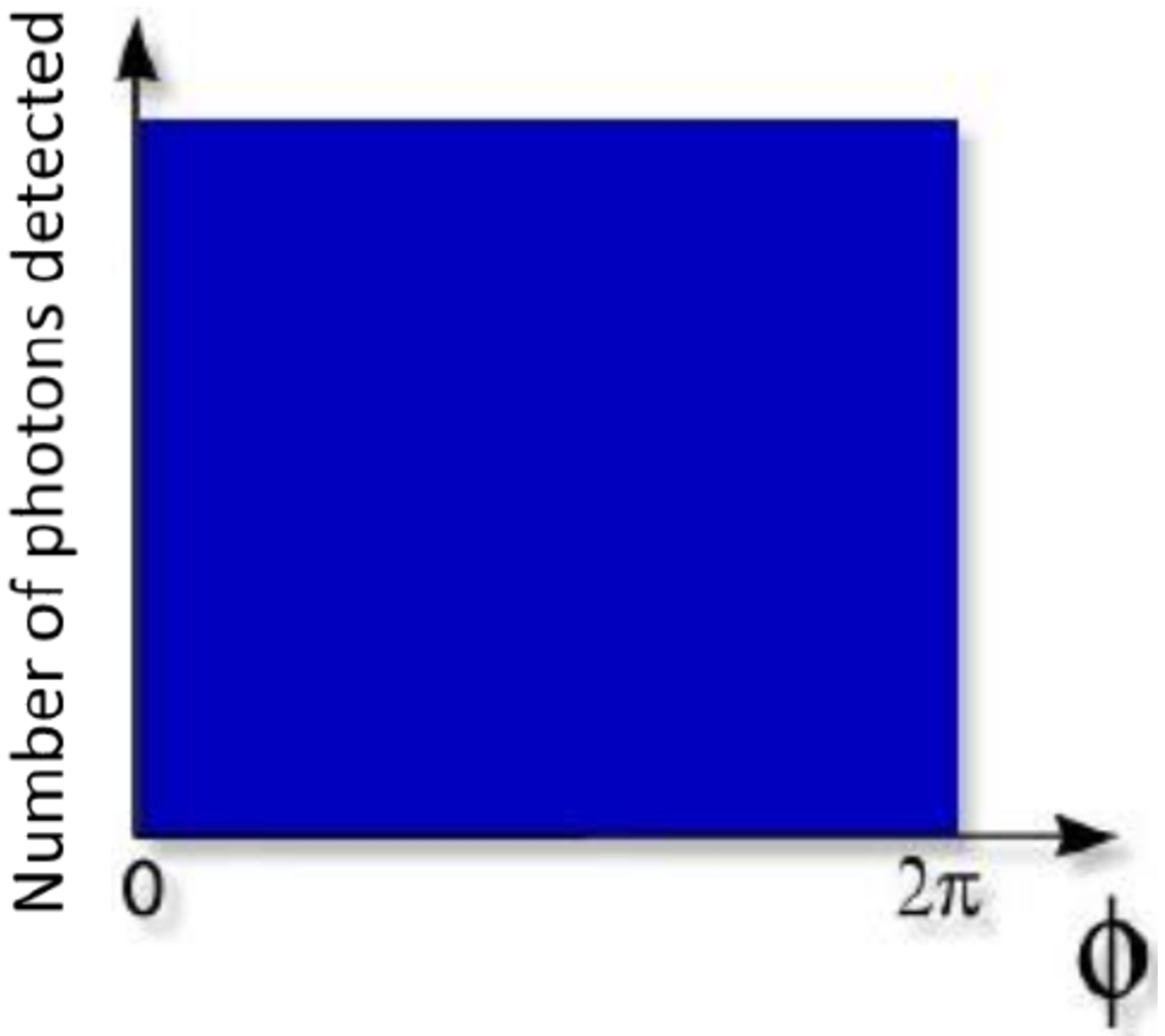}
\includegraphics[width=5 cm]{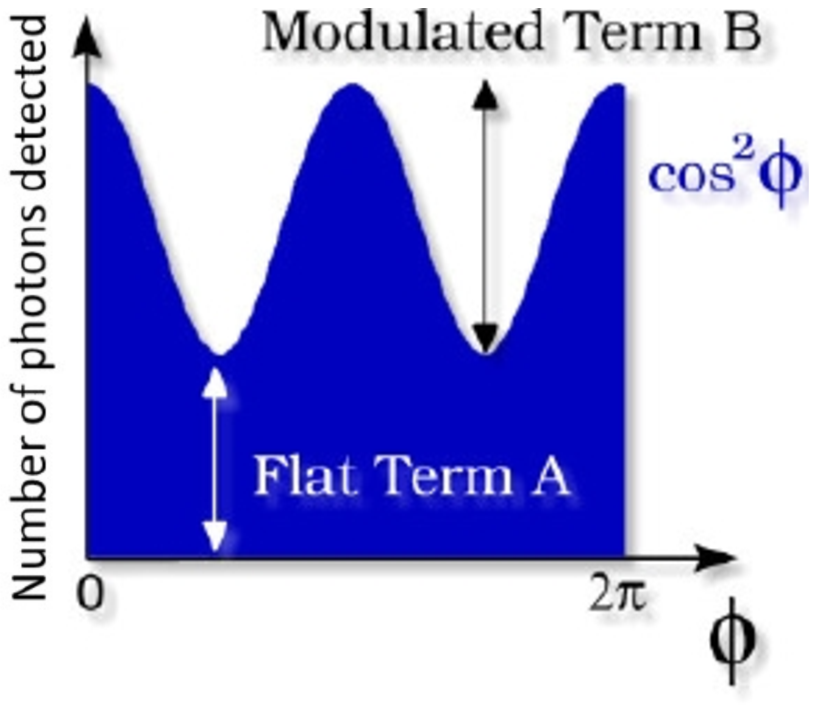}
\caption{On the left panel: General view of the response of a polarimeter. The detector response to unpolarised radiation is \textit{flat} since all the angular directions have the same probability. On the right panel: The detector response to the polarized radiation is modulated since one angular direction is more probable.}
\label{fig:flatmod}
\end{figure}

A polarimeter is a detector that analyses different angular directions and detects photons with respect to these directions. The current technology available for X-ray polarimetry allows to deal with the linear polarization only. If the radiation is not polarized every angular direction has the same probability, thus the number of photons detected, as a function of the angular directions, is the same and the detector response is \textit{flat} (see left panel of Figure~\ref{fig:flatmod}). If the radiation is polarized one angular direction will be more probable and a $\cos^2$ modulation arises (see right panel of Figure~\ref{fig:flatmod}). The $\cos^2$ modulation depends on the dipole interaction between the photon and the interacting electron of an atom in the sensitive volume of the polarimeter. The modulation function can be defined as in Equation~(\ref{eq:modfunction1}), where $A$ is the flat term and $B$ is the modulated term:
\begin{equation}
N(\phi)=A_P+B_P\cos^2(\phi-\phi_0) \label{eq:modfunction1}
\end{equation}

The modulated fraction of the detector response is given by the modulation factor $\mu$ that is defined for $100\%$ polarized radiation:
\begin{equation}
\mu=\dfrac{N_{100\%}^{\mathrm{max}}-N_{100\%}^{\mathrm{min}}}{N_{100\%}^{\mathrm{max}}+N_{100\%}^{\mathrm{min}}}=\dfrac{B_{100\%}}{2A_{100\%}+B_{100\%}}
\label{eq:modulationfactor}
\end{equation}
where $N_{100\%}^{\mathrm{max}}$ and $N_{100\%}^{\mathrm{min}}$ are the maximum and minimum number of photons detected in the angular bins of the modulation histogram for $100\%$ polarized radiation.
It is a normalization term needed to derive the polarization degree $P$ for radiation with an unknown polarization:
\begin{equation}
P=\dfrac{1}{\mu}\dfrac{B_{P}}{2A_{P}+B_{P}}
\label{eq:polarization1}
\end{equation}

The performance of different polarimeters can be compared by means of the Minimum Detectable Polarization (MDP), which is defined as the minimum polarization that can be detected at a confidence level of 99$\%$ \citep{Weisskopf2010,Strohmayer2013}:
\begin{equation}
\mathrm{MDP(99\%)}=\dfrac{4.29}{\mu R}\sqrt{\dfrac{R+B}{T}}\label{eq:mdp}
\end{equation}

In this equation $R$ is the rate of the source, $B$ is the background and $T$ is the observation time. A~signal, with a modulation corresponding to a polarization lower than the MDP, is compatible with a statistical fluctuation and therefore, no positive detection of a polarization can be claimed.

With some trigonometry it is possible to find a more convenient formula to fit the modulation curve on the histogram of the angular directions with respect to Equation~(\ref{eq:modfunction1}). It is given by \citep{Kislat2015}:
\begin{equation}
N(\phi)=N_0\cdot(1+a\cos(2(\phi-\phi_0)))
\label{eq:}
\end{equation}
where the modulation factor is directly derived. As in Equation~(\ref{eq:polarization1}) the polarization degree is given by:
\begin{equation}
P=\dfrac{a}{\mu}
\label{eq:polarizaiton2}
\end{equation}

The quality factor $Q$ is a parameter useful to compare the sensitivity of different polarimeters. It~is derived from the MDP (see Equation~(\ref{eq:mdp})) by assuming a source dominated observation ($B\simeq0$):
\begin{equation}
\mathrm{MDP(99\%)}\propto \dfrac{1}{\mu \sqrt{\epsilon}}\dfrac{1}{\sqrt{F}}=\dfrac{1}{Q}\dfrac{1}{\sqrt{F}}
\label{eq:MDPqfactor}
\end{equation}
where $\epsilon$ is the detector quantum efficiency and $F$ is the source flux.
The quality factor is:
\begin{equation}
Q=\mu \sqrt{\epsilon}
\label{eq:qualityfactor}
\end{equation}

Obviously, the formalism introduced so far is compatible with Stokes parameters, which are typically used for those wavelengths for which the intensity of the radiation is used instead of single photon counting.
A comprehensive treatment of the relation between the typical high energy formalism and the Stokes parameters is given in \citep{Strohmayer2013, Kislat2015}.

\section{Polarimeters and Systematics}\label{sec:systematics}

The detector geometry is crucial for polarimetry, more than for other observational techniques. A~not~well conceived geometry can originate large systematics, that need to be controlled. For example, a hexagonal geometry of the sensitive elements (pixels or scintillating bars) is preferable with respect to a squared geometry, due to the $\pi/2$ spurious modulation introduced.
The rotation of the polarimeter, with respect to the line of sight, may be used to control this effect. However, to rotate the detector~is~not always possible in the mission design.
The scattering and photoelectric polarimeters, the signals~of which depend on the azimuthal response, show a spurious modulation if the incoming beam of radiation is inclined with respect to the detector axis (the azimuthal
symmetry is broken). This effect~is more relevant for larger inclinations and it is not negligible for an inclination of some degrees. Typically, a correction is applied by comparing the off-axis modulation curve with
the on-axis one \citep{Jourdain2012}. However, a theoretical treatment of this effect, with the correction to apply, is described in \citep{Muleri2014}.

\section{Polarimeter Techniques and Instrumentation}\label{sec:techniques}

In the X-rays it is possible to identify three different energy bands in which different physical processes can be exploited to perform polarimetry (see Table~\ref{tab:poltechniquesscigoalstab}). The diffraction on crystals, especially in the past, and currently on multilayer mirrors, are used below 1 keV. At higher energies, up to some tens of keV, the photoelectric effect is exploited nowadays, while Thomson scattering was previously used. The Compton scattering instead is exploited in the 100 keV energy range.

\begin{table}[H]
\caption{Polarimetry techniques and scientific goals for different energy bands. \label{tab:poltechniquesscigoalstab}}
\centering
\tablesize{\footnotesize}
\begin{tabular}{p{2.5cm}p{2.2cm}p{2.5cm}p{2.5cm}p{2.7cm}}
 \toprule
\multicolumn{1}{c}{ }& &\textbf{< 1 keV} &\textbf{1--10 keV}& \textbf{> 10 keV} \\
\midrule

 \multirow{6}{*}{\textbf{Polarimetry~techniques}} &		&\multirow{6}{*}{$\bullet$Diffraction} &\vspace{6pt} $\bullet$Photoelectric effect, $\bullet$Thomson~scattering (from few keV) & $\bullet$Photoelectric effect (up to tens of keV), \hspace{3cm }$\bullet$Thomson~scattering (up to tens of keV), \hspace{3cm} $\bullet$Compton~scattering (from few tens of keV)\\
\midrule

\textbf{Scientific goal}	&\textbf{Sources}	& \multicolumn{3}{c}{} \\ \midrule

\multirow{8}{4em}{Acceleration phenomena}  & \multirow{2}{*}{PWN} 	& yes \hspace{3cm} (but absorption)	& \multirow{2}{*}{yes}	 & \multirow{2}{*}{yes} \\
&  SNR 	& no	& 	yes & yes \\
& \multirow{2}{*}{Jet (microquasars)}	& yes \hspace{3cm} (but absorption)	& \multirow{2}{*}{yes}	 & \multirow{2}{*}{yes} \\
& Jet (blazars)	& yes	& yes	 & yes \\
& \multirow{2}{*}{Solar flares}	& difficult \hspace{3cm} (but thermal~\&~lines)	& difficult \hspace{3cm}  (but thermal~\&~lines)	& \multirow{2}{*}{yes} \\ \midrule
\multirow{7}{4em}{Emission~in~strong magnetic~field} & \multirow{2}{*}{WD}	& yes  \hspace{3cm} (but absorption)	& \multirow{2}{*}{yes}	 & \multirow{2}{*}{difficult} \\
 & AMs	& no  & yes	 & yes \\
  & \multirow{2}{*}{X-ray pulsator}	& \multirow{2}{*}{difficult}  & yes  \hspace{3cm}  (no cyclotron?)	 & \multirow{2}{*}{yes} \\
   & \multirow{2}{*}{Magnetar}	& yes  \hspace{3cm}  (better)  & \multirow{2}{*}{yes}	 & \multirow{2}{*}{no} \\ \midrule
\multirow{4}{4em}{Scattering~in aspherical~geometries} & Corona in XRB and~AGN	& \multirow{2}{*}{difficult}  	& \multirow{2}{*}{yes}	 & yes   \hspace{3cm} (difficult) \\
 & X-ray   \hspace{3cm}  reflection nebulae & \multirow{2}{*}{no}  	& yes   \hspace{3cm}  (long exposure)	 & \multirow{2}{*}{yes} \\  \midrule

\multirow{7}{4em}{Fundamental~physics }& \multirow{2}{*}{QED (magnetar)}	& yes   \hspace{3cm}   (better)  	& \multirow{2}{*}{yes}	 &\multirow{2}{*}{ no} \\
 & GR (BH) & no  	& yes    & no \\
  & QG (blazars) & difficult  	& yes    & yes \\
   & Axions \hspace{3cm} (blazars, clusters) & \multirow{2}{*}{yes?}  	& \multirow{2}{*}{yes}   & \multirow{2}{*}{difficult} \\

\bottomrule
\end{tabular}
\end{table}

\subsection{Bragg Diffraction Polarimeters}

In a crystal, with lattice spacing $d$, diffraction occurs for an energy $E$ at an angle $\theta$, if the following equation is verified:
\begin{equation}
E=\dfrac{nhc}{2d\sin(\theta)}
\label{eq:braggformula}
\end{equation}
where $n$ is the diffraction order, $h$ is the Planck constant and $c$ is the speed of light.
This formula has~to~be verified in a very narrow energy band (few eV), thus mosaic crystals or bent crystals assemblies are typically used to increase the angle of acceptance and, therefore, the energy band.
The~incoming radiation beam can be thought as comprising two components: the $\pi$ component, which~lies parallel to the incidence plane, and the $\sigma$ component, which is perpendicular to this plane.

\begin{figure}[H]
\centering
\includegraphics[width=11cm]{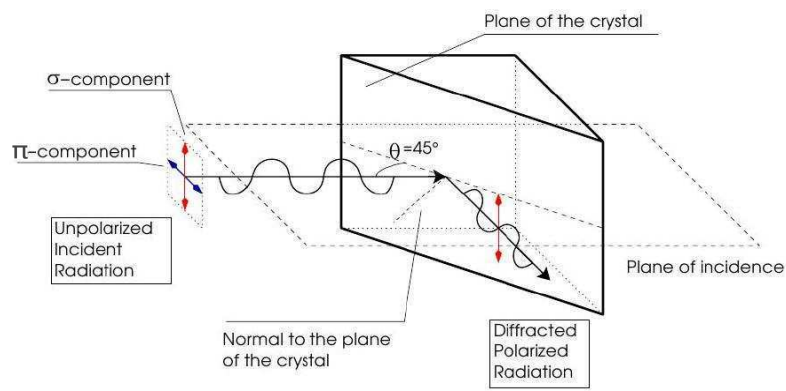}
\caption{Schematic representation of Bragg diffraction at 45$^{\circ}$ on a crystal. The radiation component, which is polarized perpendicularly with respect to the plane of incidence, is efficiently diffracted, while~the radiation component that have a parallel polarization with respect to this plane is absorbed by the~crystal.}
\label{fig:Bragg}
\end{figure}

If the diffraction angle is $45^{\circ}$ only the $\sigma$ component survives and the out-coming beam is $100\%$ polarized, orthogonally with respect to the incidence plane, as shown in Figure~\ref{fig:Bragg}.
By rotating the crystal around the beam axis, the flux of the out-coming radiation is modulated, since the polarized component is alternatively the $\sigma$ and $\pi$ component (the modulation period is twice the rotation period).
However, the Bragg formula (see Equation~(\ref{eq:braggformula})) has to be verified to diffract the polarized component. Therefore, the spacing $d$ is chosen such that the incidence angle $\theta$, for the specific energy, is close as possible to $45^{\circ}$.
This technique is extremely inefficient to measure the polarization of continuous energy spectra, but allows to analyse photons at energies starting from less than 1 keV. A~detailed description~of the Bragg diffraction on crystals, for polarimetry, is given in \cite{Silver1989}.
Currently, a~large~technological effort is focused in developing multilayer mirrors to exploit the Bragg diffraction. Multilayer
mirrors allow, with respect to crystals: (1) to choose, to some extent, the energy of interest; (2) to shape the mirror as a paraboloid and to focus the incident beam, thus to maximize the signal to noise ratio. 

In Tab.~\ref{tab:diffrpoltab} relevant Bragg diffraction polarimeters and missions planned are listed.
\subsubsection{SOLPEX: SOLar Spectroscopy and Polarimetry EXperiments for Solar Flares}

Hard X-rays are well suited to perform polarimetry of solar flares, since at energy larger than about 15--20~keV, a highly polarized component of non-thermal Bremsstrahlung \citep{Zharkova2011} is dominant in the energy spectra, with respect to the thermal one \citep{Emslie1980a}. Moreover, at energy lower than 10~keV, there is a large component of emission lines from hot highly ionized elements \citep{Doschek2002}. However, the non-thermal component can be large even at low energy in the early impulsive phase of flares.

The low energy band is the target of B-POL (Bragg POLarimeter) in the SOLPEX experiment. It~will~be mounted on the KORTES platform, on board the Nauka Russian module, on the International Space Station (ISS) \cite{Steslicki2015}. The launch of the KORTES platform to the ISS is expected in the period 2022--2023 \cite{Vishnyakov2017}.
B-POL comprises a Si(111) bent crystal (85.5 mm $\times$ 31 mm) at the Brewster angle~of~$\sim$45$^\circ$. The bending radius of 610.0~mm allows a spectral range of 3.940--4.505~\AA. The readout detector is a Charge Coupled Device (CCD). The polarimeter system rotates at 1 rev./s to ensure the measurement of the polarization.
The ISS environment allows an easy access and maintenance and essentially no power limitation. However, the inclined orbit causes an eclipse every 90~min. Moreover, due to the ISS motion, only few minutes of uninterrupted observation per orbit are possible.

\subsubsection{LAMP: Lightweight Asymmetry and Magnetism Probe}

Multilayer mirrors were proposed in the early 2000s for the PLEXAS polarimeter \citep{Marshall2003} to perform polarimetry by exploiting the Bragg diffraction and focusing soft X-rays (250 eV). Recently, this idea had a new boost with the project of the LAMP polarimeter. It is a micro-satellite mission concept~for astronomical X-ray polarimetry, which is currently under an early assessment phase in~China. The~scientific goals comprise the study of the thermal radiation from the surface of pulsars and the synchrotron radiation produced by relativistic jets in blazars.
X-ray photons at 250~eV are focused by a segmented paraboloidal multilayer mirror with a collecting area of about 1300~cm$^2$. A~position sensitive detector is placed at the focal plane \cite{She2015}.
The reflection angle varies along the mirror and the thickness changes to match the Bragg law for the 250~eV photons anywhere on the mirror. The focal plane detectors, that match the LAMP goals, are both a CCD or the Gas Pixel Detector (GPD) \cite{Costa2001, Bellazzini2007} used as a simple imaging detector. Because of the low-cost mission profile based on a micro-satellite, the GPD is preferable with respect to the CCD.

\subsubsection{REDSoX: Rocket Experiment Demonstration of a Soft X-Ray Polarimeter}

The REDSoX polarimeter \cite{Marshall2017,Marshall2018} is a demonstration experiment aimed to perform polarimetry by means of multilayer laterally graded mirrors \citep{Panini2018}. The~scientific goals of the REDSoX polarimeter comprise the measurement of the polarization of pulsars, AGNs, jets in binary systems and discs.
The~X-ray polarization is measured by means of a chain which comprises critical angle transmission (CATs) gratings that disperses the radiation by matching the Bragg condition at the first order of laterally graded multilayer mirrors (LGMLs) that illuminate CCD detectors.

\begin{table}[H]
\caption{Bragg diffraction polarimeter experiments and missions planned. \label{tab:diffrpoltab}} 
\centering
\scalebox{0.92}[0.95]{\begin{tabular}{cccp{3.2cm}cp{2.8cm}c}
\toprule
\textbf{Name}	&\textbf{Time Schedule} &\textbf{Focal Plane}& \textbf{F.O.V} & \textbf{En. Range}& \textbf{Science Obj.}& \textbf{References}\\
\midrule
 SOLPEX	& launch 2022--2023	& no	 &  $2 \times 2$~arcmin$^2$	& 3.940--4.505~\AA & solar flares & \cite{Vishnyakov2017,Steslicki2015}\\ \midrule
\multirow{2}{*}{ LAMP}	& \multirow{2}{*}{assesment}	& \multirow{2}{*}{yes}	 & \multirow{2}{*}{few arcmin}  &\multirow{2}{*}{250 eV}& pulsars~thermal~emission, jet~blazars	&  \multirow{2}{*}{\cite{She2015}}\\ \midrule
 \multirow{3}{*}{REDSoX}	& \multirow{3}{*}{development}	& \multirow{3}{*}{yes}	 &  \multirow{3}{*}{pointing jitter $\lesssim$ 15 arcsec} &\multirow{3}{*}{0.2--0.8 keV}& AGNs and~binaries jets and discs, Isolated NSs and pulsars	& \multirow{3}{*}{\cite{Marshall2017,Marshall2018,Panini2018}} \\
\bottomrule
\end{tabular}}

\end{table}

\subsection{Photoelectric Polarimeters}

\begin{figure}[H]
\centering
\includegraphics[width=8 cm]{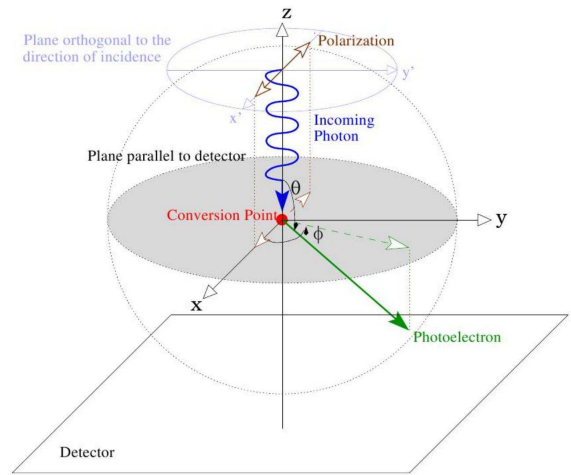}
\caption{Scheme of the photoelectric effect absorption. See Equation~(\ref{eq:phcrosssection}).}
\label{fig:phscheme}
\end{figure}

When a photon is absorbed via photoelectric effect by an atom, a photoelectron is emitted. The~direction~of emission depends on the differential cross-section defined as \citep{Ghosh1983}:
\begin{equation}
\dfrac{d\sigma_{ph}}{d\Omega}=\dfrac{\sigma_{ph}^{\mathrm{tot}}}{4\pi}\left[  1+\dfrac{b}{2} \left(  \dfrac{3\sin^2(\theta)\cos^2(\phi)}{(1+\beta \cos(\theta))^4}-1 \right)  \right]
\label{eq:phcrosssection}
\end{equation}
where $\theta$ and $\phi$ are the photoelectron polar and azimuthal angles of emission, respectively (see~Figure~\ref{fig:phscheme}). The electric vector of the absorbed photon defines the $\phi=0$ angular direction in Figure~\ref{fig:phscheme}. The~photoelectron is emitted with a higher probability parallel to this direction.
The parameter $b$ is called orbital asymmetry factor and it is equal to 2 for $s$ orbitals, while it is less than 2 for the other cases. In the first case, the constant term vanishes and only the $cos^2(\phi)$ modulated term remains. Therefore, the photo-absorption due to spherical orbitals, is a perfect polarization analyser.
A gas detector is suitable to perform polarimetry by exploiting the photoelectric effect, because the photoelectron track length in gas is in the range of millimetres for absorbed photons of energy from 1 keV up to some tens~of~keV.

\begin{figure}[H]
\centering
\includegraphics[width=8 cm]{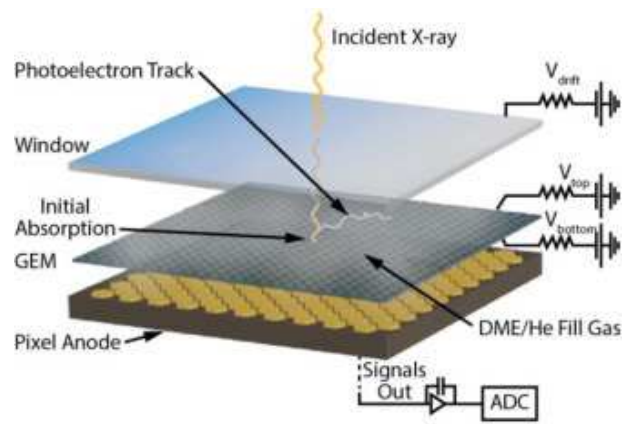}
\caption{The GPD polarimeter concept design.}
\label{fig:GPD}
\end{figure}

\begin{figure}[H]
\centering
\includegraphics[width=8 cm]{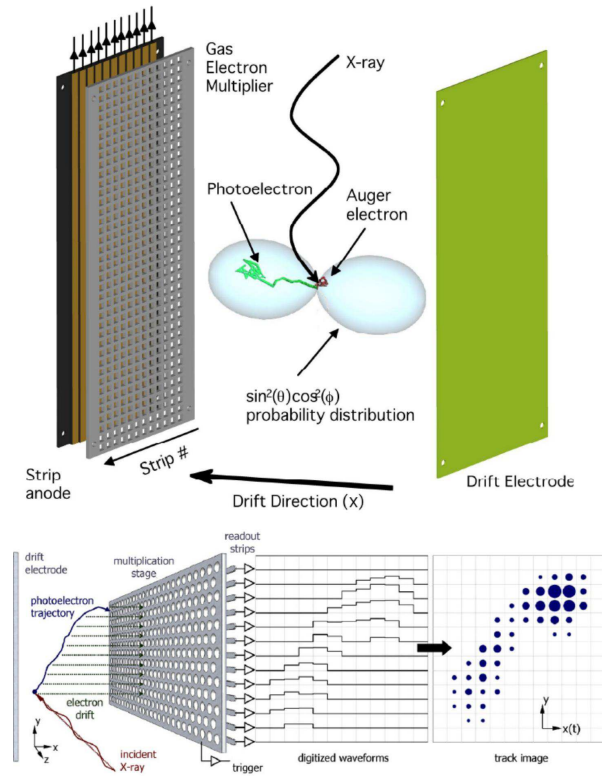}
\caption{The TPC polarimeter concept design.}
\label{fig:TPC}
\end{figure}

Two different technologies are available to perform polarimetry by exploiting the photoelectric effect. The schematic views of these two kinds of detectors are shown in Figures~\ref{fig:GPD} and~\ref{fig:TPC}. They are the Gas Pixel Detector (GPD) \citep{Costa2001, Bellazzini2007} (developed in Italy by INFN and INAF-IAPS research institutes) and the Time Projection Chamber (TPC) \citep{Black2007} (developed in USA by the GSFC). In both cases, the X-ray photon is absorbed in a gas cell and the ionization charges, produced by the photoelectron, are drifted and multiplied by a Gas Electron Multiplier (GEM) and, eventually, read out by an anodic plane.

In the GPD the readout plane is a finely subdivided pixel plane, while in the TPC the readout is performed by means of a strip detector. The other relevant difference between these two polarimeters is given by the orientation of the readout plane with respect to the optical axis. In the GPD the incoming photon direction is perpendicular~to the readout plane. Therefore, the readout plane is also parallel~to the plane of maximum probability of emission of the photoelectrons ($\theta=\pi/2$ in Equation~(\ref{eq:phcrosssection})). This~geometry allows to make the image of the tracks along their extension and the ionization charges are affected by the same drift field (given the angle $\theta$) for all the $\phi$ angles from 0 to $2\pi$.

Differently with respect to the GPD, in the TPC the incoming photon direction is parallel to the readout plane. The ionization charges are drifted orthogonally with respect to the optical axis and the detector quantum efficiency can be increased simply by making a deeper gas cell. Moreover, its length does not affect the charge diffusion along the drift. On the contrary, in the GPD, a thicker gas cell implies a larger diffusion and ionization charge dispersion. However, the TPC geometry is intrinsically asymmetric, since for $\theta\sim\pi/2$ (see Equation~(\ref{eq:phcrosssection})), photoelectrons are emitted towards the GEM or in the opposite direction. This asymmetry produces systematics which can to be mitigated by the rotation~of the polarimeter and by a careful timing analysis on the track readout signals (ionization charges produced closer to the GEM give rise to a signal earlier with respect to the other ones as shown in the bottom panel of Figure~\ref{fig:TPC}).

The TPC design was proposed on board GEMS \citep{Swank2010,Jahoda2010,Hill2012,Enoto2014,Kitaguchi2014} and PRAXyS \citep{Hill2016,Iwakiri2016,Tamagawa2017,Nakano2017} mission proposals.
The GPD design was proposed in the past in many mission proposals (the most relevant are: IXO~\citep{Bellazzini2010b}, NHXM \citep{Tagliaferri2010}, POLARIX \citep{Costa2010}, ESA Medium class mission XIPE \citep{Soffitta2016}) and currently in the IXPE mission~\cite{Weisskopf2016, Soffitta2017} approved by NASA and eXTP mission \citep{Zhang2016} selected as \textit{background mission}\footnote{A background mission in the ``Strategic Priority Space Science Program'' of the Chinese Academy of Sciences is a project selected to be launched that undergoes an assessment phase. The launch time schedule is adjusted depending on the level~of technology readiness reached by the project during the assessment.} by the Chinese Academy of Sciences.

In Tab.~\ref{tab:phepoltab} the missions scheduled for a launch which host photoelectric X-ray polarimeters are listed.

\subsubsection{IXPE: Imaging X-Ray Photoelectric Polarimeter}

The GPD is currently the polarimeter on board the IXPE mission, selected by NASA in the framework of the Explorer missions, which is scheduled for the launch in 2021 \citep{Weisskopf2016,Soffitta2017}.
IXPE is a joint mission between NASA and ASI. The focal plane comprises three GPD polarimeters provided by the Italian collaboration.
The GPD on board IXPE allows to perform imaging spectro-polarimetry due to the capability to measure the interaction point, the ionization charge (energy resolution $\dfrac{\Delta E_{FWHM}}{E}\simeq 16\%$ at 5.9~keV) and the direction of emission of the photoelectrons ($\phi$ azimuthal angles). This is obtained by means of the analysis of the statistical momenta of the ionization charge distribution of each track projected onto the readout plane. A detailed description of the current state of development of the GPD for the IXPE mission is given by \cite{Sgro2017,Muleri2017,Fabiani2017}.
The three GPDs on board IXPE are sensitive in the 2--10~keV energy range and they are filled with a gas mixture which comprises 20\% He and 80\% Dimethyl Ether (DME) at 1 bar of pressure in a 1 cm thick gas cell. The active area is $15\times 15 $~cm$^2$ and it is subdivided in $300\times 352$ pixels with 50 $\upmu$m of pitch. The total effective area of the three telescopes is 854~cm$^2$ at 3~keV, the angular resolution is $\leq$30~arcsec and the field of view is $12.8\times 12.8$~arcmin.
IXPE will conduct precise X-ray polarimetry for several categories of cosmic sources, ranging from neutron stars and stellar-mass black holes, to SNRs and AGNs. For the brighter and extended sources, such as PWNe, SNR and AGN large scale jets, IXPE will perform X-ray polarimetric imaging for the first time.
For~example, the imaging capability will allow to obtain the polarimetric map, with $\leq$30~arcsec angular resolution, of the Crab nebula, improving the result obtained in the '70 by OSO~8, which measured the mean polarization of the entire source \citep{Weisskopf1978}.

\subsubsection{eXTP: Enhanced X-Ray Timing and Polarimetry Mission}

The GPD is also the polarimetric detector on board the eXTP mission \citep{Zhang2016}. The launch is planned earlier than 2025\footnote{\url{http://www.isdc.unige.ch/extp/}.}.
The primary goals of eXTP comprise the study of the equation of state of matter at supra-nuclear density, Quantum Electro-Dynamics (QED) effects in highly magnetized star and accretion in the strong-field regime of gravity. eXTP exploits simultaneous spectral-timing observations (Silicon Drift Detector (SDD)) in the 0.5--30~keV energy range and polarimetry (GPD) in the 2--10~keV energy range. eXTP is a mission selected as \textit{background mission} in the ``Strategic Priority Space Science Program'' of the Chinese Academy of Sciences since 2011. The mission consortium comprises Chinese, European and USA institutions.
In the updated configuration eXTP comprises 4 telescopes for X-ray polarimetry in the 2--10 keV energy band (GPD 20/80 He/DME at 1 bar of pressure, 1~cm thick gas~cell). The effective area of the telescopes is 500~cm$^2$ at 2 keV with a field of view of 12~arcmin and an angular resolution better than 30~arcsec.

\subsubsection{Photoelectric Polarimetry with the GPD towards Hard X-Rays}

\begin{figure}[H]
\centering
\includegraphics[width=9 cm]{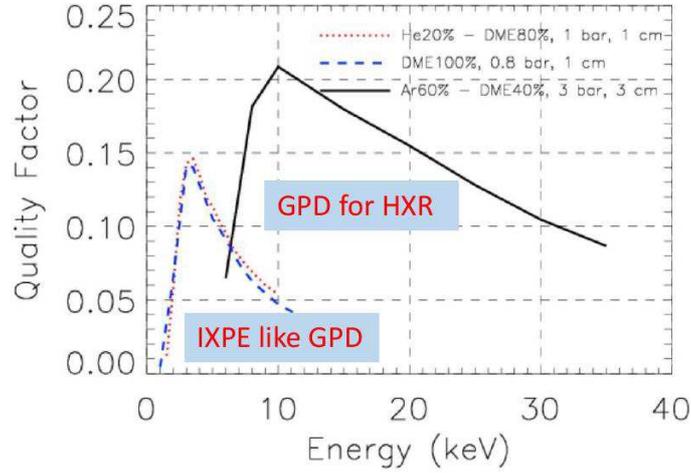}
\caption{Quality factor for different configurations of the GPD.}
\label{fig:qfactor}
\end{figure}

At energies larger than 10~keV, to increase the quantum efficiency of the GPD, a heavier gas mixture (for example using Ar in place of He) and a thicker absorption gap are needed.
In Figure~\ref{fig:qfactor}, the quality factor (see Equation~(\ref{eq:qualityfactor})) is reported for the low and high energy configurations of the GPD. The peak of the quality factor represents the peak of the sensitivity of the polarimeter. This peak is around 3~keV for a He based, or only DME, gas mixtures, at a pressure of 1 bar with an absorption gap of 1~cm. An Ar based gas mixture at 3~bar of pressure with an absorption gap of 3~cm shows the  sensitivity peak at about 10 keV \citep{Fabiani2012d}.

A hard X-ray configuration, sensitive up to some tens of keV, is suitable for solar flare polarimetry. The high flux of solar flares allows polarimetry also without optics. However, in this case no image~is obtained and only the integrated polarization of the flaring region is possible.
The hard X-ray configuration of the GPD was proposed both for distant astrophysical sources \citep{Tagliaferri2011} and the Sun~\citep{Berrilli2015,Fabiani2013a}, but never approved.

\begin{table}[H]
\caption{Photoelectric X-ray polarimetric missions planned. \label{tab:phepoltab}} 
\centering
\begin{tabular}{cccccp{2.5 cm}c}
\toprule
\textbf{Name}	&\textbf{Time Schedule} &\textbf{Focal Plane}& \textbf{F.O.V} & \textbf{En. Range}& \textbf{Science Obj.}& \textbf{References}\\
\midrule
\multirow{3}{*}{IXPE}	& \multirow{3}{*}{launch 2021} & \multirow{3}{*}{yes}	 & \multirow{3}{*}{$12.8\times 12.8$~arcmin$^2$}	& \multirow{3}{*}{2--10 keV} & BH, NSs isolated and~accreting,~AGNs, PWNs, SNRs, Sgr~B2  & \multirow{3}{*}{\citep{Weisskopf2016,Soffitta2017,Sgro2017,Muleri2017,Fabiani2017}}\\ \midrule
 \multirow{3}{*}{eXTP}	& \multirow{3}{*}{launch before 2025}	& \multirow{3}{*}{yes}	 & \multirow{3}{*}{$12\times 12$~arcmin$^2$}   &\multirow{3}{*}{2--10 keV} & BH, NSs isolated and~accreting,~AGNs, PWNs, SNRs, Sgr~B2  	& \multirow{3}{*}{\citep{Zhang2016}}\\

\bottomrule
\end{tabular}
\end{table}

\subsection{Scattering Polarimetry}

The scattering of a photon of energy $E$ on a free electron is described by the Klein-Nishina cross section \citep{Heitler1954}:
\begin{equation}
\left( \dfrac{d\sigma}{d\Omega}(E, E^\prime, \theta, \phi) \right)_{\mathrm{KN}}=\dfrac{r_0^2}{2}\dfrac{{E^\prime}^2}{E^2}\left[ \dfrac{E}{E^\prime} +\dfrac{E^\prime}{E} -2\sin^2\theta \cos^2\phi \right]
\label{eq:KN}
\end{equation}

The out-coming photon has an energy $E^\prime$ which is:
\begin{equation}
E'=\dfrac{E}{1+\frac{E}{m_ec^2}(1-cos\theta)}
\label{eq:EEprimo}
\end{equation}

The scattering angle $\theta$ is the angle between the direction of the incoming photon and the out-coming one. The azimuthal angle $\phi$ is $0$ for the direction parallel to the electric vector of the incoming photon.
The Equation~(\ref{eq:KN}) tells us that: (1) the maximum of the $\cos^2$ term is for $\phi=0$, due to the minus sign of the modulated term. Therefore, differently from the photoelectric effect, the photon is scattered with a higher probability orthogonally to the direction of polarization of the incoming photon; (2) The constant term $\dfrac{E}{E^\prime} +\dfrac{E^\prime}{E} $ makes the incoherent scattering not a perfect polarization analyser. Only if $E=E^\prime$ (Thomson limit), for scattering at $\theta=\pi/2$ the Klein-Nishina cross section is re-written as the Thomson cross section for scattering at $\theta=\pi/2$:
\begin{equation}
\left( \dfrac{d\sigma}{d\Omega} (\frac{\pi}{2}, \phi)  \right)_{\mathrm{Th.}}=r_0^2\sin^2 \phi
\label{eq:Thomson}
\end{equation}
which describes a perfect polarization analyser.

The theoretical modulation factor achievable by an ideal Compton polarimeter is:
\begin{equation}
\mu(\theta)=\frac{N_\mathrm{max}(\theta)-N_\mathrm{min}(\theta)}{N_\mathrm{max}(\theta)+N_\mathrm{min}(\theta)}=\frac{(\frac{d\sigma}{d\Omega})_{\phi=\frac{\pi}{2}}-(\frac{d\sigma}{d\Omega})_{\phi =0}}{(\frac{d\sigma}{d\Omega})_{\phi=\frac{\pi}{2}}+(\frac{d\sigma}{d\Omega})_{\phi =0}}=\frac{\sin^2\theta }{\frac{E}{E^\prime}+\frac{E^\prime}{E}-\sin^2 \theta} \label{eq:Muphi}
\end{equation}

\begin{figure}[H]
\centering
\includegraphics[width=9 cm]{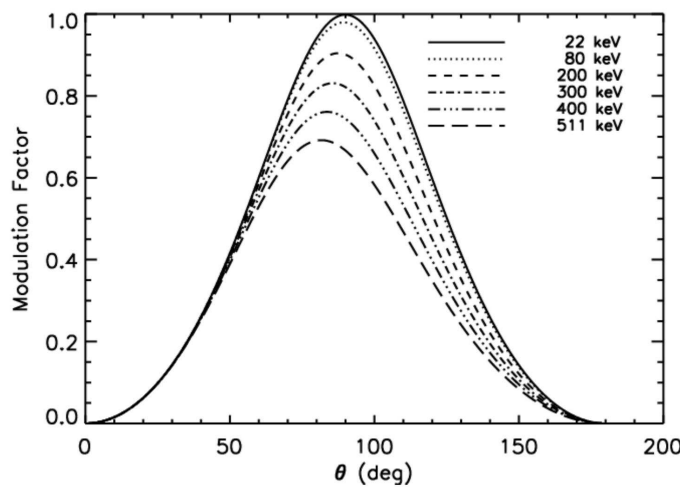}
\caption{Modulation factor as a function of the scattering angle for different energies of the incoming~photon.}
\label{fig:comptonmodfactor}
\end{figure}

The modulation factor, as expressed by Equation~(\ref{eq:Muphi}), is shown for different energies in Figure~\ref{fig:comptonmodfactor}. The modulation factor is higher for scattering at $\theta=\pi/2$ and it is lower for higher energy of the incoming radiation. At the Thomson limit, the modulation factor for scattering at $\theta=\pi/2$ is equal to 1 (perfect polarization analyser).
In Compton scattering, the interaction of the photon depends also on the influence of the atomic electrons distribution and binding energies, described by the scattering function $S(\chi,Z)$. The product of the scattering function with the Klein-Nishina cross-section gives the angular distribution of scattered photons in matter \citep{Hubbel1975}, which is:
\begin{equation}
\frac{d\sigma}{d\Omega}=\biggl(\frac{d\sigma}{d\Omega}\biggr)_\mathrm{KN} \cdot S(\chi,Z)\label{eq:KNSF}
\end{equation}
where $Z$ is the atomic number and $\chi=\sin(\frac{\theta}{2})/\lambda$[\AA], in which $\lambda$ is the incident photon wavelength.
The~scattering function essentially suppresses forward scattering with respect to the Klein-Nishina~formula.

\begin{figure}[H]
\centering
\includegraphics[width=9 cm]{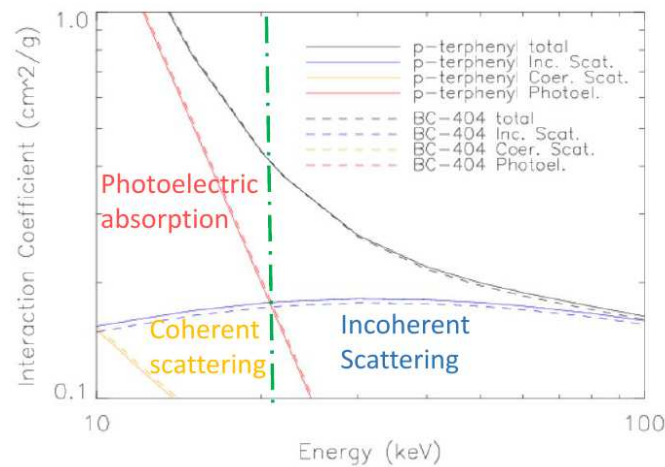}
\caption{Mass attenuation coefficients of photoelectric absorption, coherent scattering and incoherent scattering for typical plastic scintillator materials (BC-404 and p-terphenyl) which the scatterers are made of.}
\label{fig:scattcrossection}
\end{figure}

There is a wide \textit{zoology} of polarimeters based on Compton and Thomson scattering. The~incoherent nature of Compton scattering allows to deposit a certain amount of energy in the scatterer element, which can be readout in coincidence with the energy deposit in the absorber, where the scattered photon stops. This approach allows to reduce drastically the background with respect to Thomson scattering polarimeters, which do not allow to readout the energy deposition in the scatterer.
In~both~cases the scattering process competes with photo-absorption and, therefore, scatterers are made of light elements, which are scintillating (plastics or crystals) in Compton polarimeters and passive (for~example~Lithium) in Thomson polarimeters.
The incoherent scattering, in Compton polarimeters, identifies a \textit{natural} energy threshold above which the scattering process is more efficient with respect to absorption. This~is~shown in Figure~\ref{fig:scattcrossection} for two typical materials of scatterers (BC-404 and p-terphenyl), where the mass attenuation coefficint for scattering (coherent and incoherent) and photo-absorption are reported. Above about 20 keV (point dashed green vertical line)the incoherent scattering in plastic scintillators is the dominating interaction.

The Compton polarimeters can be classified into two categories, depending on the material of the scatterer and of the absorber. The \textit{one phase} polarimeters are those in which the scatterer and the absorber are made of the same low $Z$ material. The \textit{two phase} polarimeters are those in which the scatterer is made of a low $Z$ material (higher scattering probability with respect to photoelectric
absorption) and the absorber is made of a high $Z$ material, which maximize the absorption
probability.
Moreover, scattering polarimeters can be focal plane detectors for pointed observations (large effective area depends on the optics) or non-focal plane ones, which can have a large field of view (for~example~to observe GRBs) or a narrow field of view for pointed observations, if they are collimated.

The typical critical parameters for scattering polarimeters are:
\begin{itemize}[leftmargin=*,labelsep=5.8mm]
\item background rejection, if large sensitive volumes are involved. The mitigation of background is possible by means of the scatterer/absorber coincidence (it is
intrinsic of the Compton scattering technique), anticoincidence, passive
shielding and a careful estimation of sensitive volumes needed.
For Thomson polarimeters background is a very critical issue.
\item Scintillation light cross-talk. The mitigation of this effect is possible by means of a careful choice and application of the wrapping around the scintillating elements.
\item Scintillating element light loss (for example from the edges).
The mitigation of this effect is possible by means of a careful choice of the wrapping and of the optical
contact between the interfaces towards the light sensor.
\end{itemize}

Many scattering polarimeters have been proposed and a lot of demonstrators and pathfinder are flown until nowadays. Relevant examples are discussed in the following sections and their characteristics are sumarized in Table~\ref{tab:scatpoltab}.

\begin{table}[H]
\caption{Relevant scattering polarimeters examples. \label{tab:scatpoltab}}
\centering
\scalebox{0.95}[0.95]{\begin{tabular}{p{1cm}p{1.5cm}p{2cm}p{1.5cm}p{2.2cm}cp{2cm}c}
\toprule
\textbf{ }	&\textbf{Name}	&\textbf{Time Schedule} &\textbf{Focal Plane}& \textbf{F.O.V} & \textbf{En. Range}& \textbf{Science Obj.}& \textbf{References}\\
\midrule
\textbf{Thomson}	&\textbf{ }	& \textbf{ }	& \textbf{ }& \textbf{ }& \textbf{ }& \textbf{ }& \textbf{ }\\
\midrule
& \multirow{2}{*}{POLIX}	& \multirow{2}{*}{launch 2019}& \multirow{2}{*}{no}	 & \multirow{2}{*}{$3^\circ \times 3^\circ$}	& \multirow{2}{*}{5--30~keV}& accr. powered pulsars, BH & \multirow{2}{*}{\cite{Paul2010, Paul2016}}\\ \midrule
& SPR-N	& launched 2001	& no	 & -   &20--100 keV& solar flares	& \cite{Zhitnik2006,Zhitnik2014}\\
\midrule
\textbf{Compton~1~phase}	&\textbf{ }	& \textbf{ }	& \textbf{ }& \textbf{ }& \textbf{ }& \textbf{ }& \textbf{ }\\
\midrule
& POLAR	&launched 2016	& no		& $\sim$1/3 of full sky & 50--500 keV	& GRB & \cite{Xiao2017a, Xiao2017b, Kole2018}\\ \midrule
& PoGOLite &launched 2013	& no	 &$\sim$$2^\circ \times 2^\circ$ &20--240 keV &Crab emission	& \cite{Chauvin2016, Chauvin2017, Friis2018, Mikhalev2018}\\ \midrule
& \multirow{3}{*}{PoGO+} 	& \multirow{3}{*}{launched 2016} & \multirow{3}{*}{no}	& \multirow{3}{*}{$\sim$$2^\circ \times 2^\circ$}  &\multirow{3}{*}{20--150 keV} &Crab pulsar and nebula, Cygnus X-1 	& \multirow{3}{*}{\cite{Chauvin2016, Chauvin2017, Friis2018, Mikhalev2018}}\\ \midrule
& AstroSat CZTI (imager) 	& \multirow{3}{*}{launched 2015} & \multirow{3}{*}{no}	& \multirow{3}{*}{-}  &\multirow{3}{*}{100--380 keV} &Crab~pulsar~and nebula,  bright X-ray sources & \multirow{3}{*}{\cite{Vadawale2018a, Vadawale2015}}\\
\midrule
\textbf{Compton~2~phases}&\textbf{ }	& \textbf{ }	& \textbf{ }& \textbf{ }& \textbf{ }& \textbf{ }& \textbf{ }\\
\midrule
& \multirow{2}{*}{X-Calibur}	&launched 2014, 2016	& \multirow{2}{*}{yes} & \multirow{2}{*}{8 arcmin at 20 keV} & \multirow{2}{*}{20--60 keV} & BHs,~NSs,~magnetars, AGN jets &\multirow{2}{*}{\citep{Kislat2017,Endsley2015,Beilicke2014,Beilicke2015,Guo2013}}\\ \midrule
& \multirow{3}{*}{PolariS} &\multirow{3}{*}{assesment}	& yes/no~(also~a wide field polarimeter) & \multirow{3}{*}{10 $\times$ 10 arcmin$^2$} & \multirow{3}{*}{10--80 keV}	 & SNRs,~BHs,~accretion in X-ray pulsars, GRBs	& \multirow{3}{*}{\cite{Hayashida2016}}\\ \midrule
& \multirow{3}{*}{GRAPE}	& launched 2014 and 2016;~new design~assesment & \multirow{3}{*}{no}	& \multirow{3}{*}{wide} & \multirow{3}{*}{50--500 keV}	& transient sources, GRBs, solar flares & \multirow{3}{*}{\cite{McConnell2014,McConnell2013,Connor2010,Bloser2010,Bloser2009,Kishimoto2007}}\\ \midrule
& PHENEX &launched 2006 	& no	 & 4.8$^\circ$ & 40--200 keV	& Crab Nebula & \cite{Gunji2008,Gunji2007,Kishimoto2009}\\
\midrule
& \multirow{2}{*}{GAP} 	& \multirow{2}{*}{launched 2010} & \multirow{2}{*}{no}	& \multirow{2}{*}{$\pi$ sr} & \multirow{2}{*}{50--300 keV} & GRBs,~Crab~pulsar and nebula	& \multirow{2}{*}{\cite{Yonetoku2011a,Yonetoku2011b,Yonetoku2012,Yonetoku2010}}\\ \midrule
& $\multirow{2}{*}{SPHiNX}$ & Phase-A/B1 assessment 	& \multirow{2}{*}{no}	 & \multirow{2}{*}{$\pm 60^\circ$} & \multirow{2}{*}{50--500 keV} & \multirow{2}{*}{GRBs}	& \multirow{2}{*}{\cite{Xie2018}}\\ \midrule
& PENGUIN-M 	& launched~2009,~lost& no & - & 20--150 keV & solar flares & \cite{Dergachev2009}\\ \midrule
& PING-P 	& launch~after~2025 & no & -	& 20--150 keV &solar flares & \cite{Kotov2016}\\

\bottomrule
\end{tabular}}
\end{table}
\subsubsection{POLIX}
POLIX \cite{Paul2010, Paul2016} is a Thomson scattering X-ray polarimeter of the Raman Research Institute (RRI) for a small satellite mission of ISRO. The launch is planned in 2019. The polarimeter consists of proportional counters as X-ray detectors, placed on all sides of a scattering element. A collimator restricts the field of view of the instrument to $3^\circ \times 3^\circ$. The polarimeter rotates around the viewing axis. It operates in the 5--30 keV energy band. The MDP of 2--3\%  is achieved for a 50 mCrab source with $1\times10^6$ s of exposure.

\subsubsection{SPR-N, PENGUIN-M and PING-P}

In the recent past, since 2000, only three detectors measured the X-ray polarization of solar flares: SPR-N, RHESSI and PENGUIN-M.
RHESSI \cite{SuarezGarcia2006} (launched in 2002) was designed~as~an imaging-spectrometer exploiting Compton scattering among Ge elements. It has, therefore, polarimetric capabilities. On the contrary, SPR-N and PENGUIN-M were designed specifically to measure the polarization of X-rays by means of scattering. These polarimeters, as well as the upcoming PING-P, are part of the effort along the pathway to develop sensitive polarimeters for solar~flare~physics.

The SPR-N polarimeter was launched in 2001 on board the past mission CORONAS-F. The SPR-N polarimeter measured the X-ray polarization degree in the energy ranges 20--40 keV, 40--60 keV and 60--100 keV \citep{Zhitnik2006,Zhitnik2014}. SPR-N exploits Thompson scattering on beryllium plates mounted inside a hollow hexagonal prism. Six scintillation detectors are symmetrically mounted on the prism faces around the scatterer. A phoswich detector (CsI(Na)/plastic scintillator) is used to discriminate charged particles. The geometric area of each detector is about 8 cm$^2$. The effective area ranges from $\simeq$0.3 cm$^2$ to $\simeq$1.5~cm$^2$ at 20 keV and 100 keV, respectively.

The PENGUIN-M instrument \citep{Dergachev2009}, on board the CORONAS-PHOTON mission, is~designed~to measure the degree of linear polarization of X-rays from solar flares in the energy range 20--150~keV and their X-ray spectra in the energy range 2--500 keV by means of Compton scattering. CORONAS-PHOTON was launched in 2009, but due to a failure of the spacecraft it was lost in 2010.

Therefore, PING-P \cite{Kotov2016} is the new Compton polarimeter under development. The launch is scheduled after 2025.
PING-P comprises three p-terphenyl (C18H14) scatterers and six units of CsI(Tl) crystals as absorber detectors. It will operate in the 20--150~keV energy range. Scintillating elements are coupled to photomultiplier tubes (PMTs).

\subsubsection{POLAR}
POLAR \cite{Xiao2017a,Xiao2017b, Kole2018} is a compact wide field polarimeter developed by an international collaboration~of Switzerland, China and Poland. It was launched on 15 September 2016 to be hosted on-board the China Space
Laboratory TG-2. POLAR is aimed to measure the linear polarization of hard X-rays from transient sources between 50 keV to 500 keV.
It consists of 25 identical modules comprising 64 plastic scintillator bars, which are readout by multi-anode photo multiplier tubes (MAPMTs).
POLAR energy range is optimized for the detection of the prompt emission of the gamma-ray bursts.

\subsubsection{PoGOLite and PoGO+: Polarised Gamma-Ray Observer}
PoGOLite \cite{Chauvin2016b} is a pathfinder balloon-born experiment which flew in 2013. It operated in the 20--240~keV energy range. PoGOLite design is conceived for pointed observations, since a collimator is placed in front of an array of 61 low $Z$ plastic scintillator bars.
A detector bar is 20 cm long and it~is sandwiched between two anti-coincidence components: a BGO (Bismuth~Germanate---Bi$_4$Ge$_3$O$_12$) scintillator (4~cm long) and the collimator made of an active plastic scintillator (60 cm long). The~anti-coincidence is complemented with additional 30 rods of BGO scintillators (60~cm tall) placed around the main detector cells.
The scintillators of the detector are readout by PMTs. A first stage~of background rejection is performed by means of the pulse discrimination in the detector and in the BGO veto system. A post-flight analysis allows for a more accurate background subtraction. Moreover, an additional passive neutron shield made of polyethylene surrounds the instrument.
This experience allowed to optimize the detector to design the PoGO+ polarimeter, which flew in the summer 2016~\cite{Chauvin2016, Chauvin2017, Friis2018, Mikhalev2018, Chauvin2018}.

\subsubsection{X-Calibur}

X-Calibur \citep{Kislat2017,Endsley2015,Beilicke2014,Beilicke2015,Guo2013}
is a balloon-born hard X-ray scattering polarimeter aimed to measure the polarization from BHs, NSs and AGNs. InFOC$\mu$S (International Focusing Optics Collaboration for $\mu$-Crab Sensitivity) grazing incidence X-ray mirror focuses X-rays onto a 13~mm diameter scatterer made of a plastic scintillator rod. The absorbers are an array of Cadmium-Zinc-Telluride (CZT) detectors surrounding the scattering rod.
X-Calibur was launched in 2014 and 2016 and a new launch is scheduled in 2018/2019 from McMurdo, Antarctica.

\subsubsection{PolariS}

PolariS (Polarimetry Satellite) \cite{Hayashida2016} is a proposed small mission (JAXA) designed to perform polarimetry in the 10--80~keV energy band.
The main scientific objective of PolariS mission is the hard X-ray polarimetry of bright SNRs, BHs and NSs.
Moreover, PolariS mission is aimed to measure the X-ray and $\gamma$-ray polarization of transient sources such as GRBs by means of a wide field polarimeter based on the GAP design (see Section~\ref{sect:GAP}).
The main instrument of Polaris payload comprises three hard X-ray telescopes coupled to three Compton scattering polarimeters in a focal plane configuration.
The scattering polarimeters are an optimization derived by PHENEX, which is a non-focal plane version prototype (flown on a balloon born experiment in 2006 \citep{Gunji2008}, see~Section~\ref{sec:phenex}).
The polarimeters consists of two kinds of (plastic and GSO) scintillator pillars and MAPMTs.
A coarse imaging capability (few~arcmin) is allowed by the scattering elements that are readout separately. However, the~asymmetric light cross-talk between neighbour rods introduces systematics that need~to~be~controlled.

\subsubsection{PHENEX (Polarimetry for High ENErgy X-Ray)}\label{sec:phenex}
PHENEX was launched on a balloon flight on June 2006 to observe for about
one hour the polarization of the Crab Nebula in hard X-ray band \cite{Gunji2008}. The high level of background measured (3~times the Crab signal) and
problems to the attitude control system compromised the observation.

The PHENEX polarimeter comprises 4 modules of  6$\times$6 squared plastic scintillators surrounded~by 28 CsI(Tl) elements. The CsI(Tl) scintillators are shielded by a passive graded shields made of Pb and Sn on the side and on the top. The detector comprises a Mo collimator to perform pointed observations with an opening angle limited to 4.8$^\circ$ (FWHM) in an energy range of 20--200 keV.

\subsubsection{GAP: Gamma-Ray Burst Polarimeter}\label{sect:GAP}

GAP is flying aboard the small solar power sail demonstrator IKAROS \cite{Yonetoku2011a,Yonetoku2011b,Yonetoku2012,Yonetoku2010}.
It is a compact polarimeter (17 cm of diameter and 16 cm of height). The central scatterer is a dodecagon plastic scintillator (6 cm of length) coupled with a non-position
sensitive photomultiplier tube. It~is~surrounded~by 12 CsI scintillators, which
are also coupled to PMTs to measure coincidence signals. GAP is mounted on the bottom panel of IKAROS and observes always in the anti-solar~direction.

\subsubsection{GRAPE (Gamma RAy Polarimeter Experiment)}

The GRAPE \cite{McConnell2014,McConnell2013,Connor2010,Bloser2010,Bloser2009,Kishimoto2007} polarimeter is based on Compton scattering between low-Z scatterers and high-Z absorbers in a squared array of scintillating bars.
It is a balloon-born hard X-ray polarimeter which operate between 50 keV and 500 keV, for GRBs observations. It flew during two long duration flights in 2011 and 2014. During the first flight the Crab was observed to validate the scientific capability of the polarimeter, but the background was much higher than expected. The second flight was terminated without a sufficient data collection.

An updated design of the polarimeter underwent some assessments \citep{McConnell2016} during different mission proposals to NASA (Astrophysics Mission of Opportunity (MoO) in 2012 and Small Explorer (SMEX) in 2014).

The original design of the GRAPE detector module comprised 64 optically independent scintillator elements organized in a $8\times 8$ array. The central 6~$\times$~6 array was made of low-Z plastic scintillator (5~mm~$\times$~5~mm$\times$6~cm). It was surrounded by the remaining 28 elements of high-Z inorganic scintillator, which acted as a calorimeter to detect with high efficiency the Compton scattered photons. A single detector module was readout by multi-anode photo-multiplier tube (MAPMT).
This design was affected by a relevant optical cross-talk between adjacent elements of the scintillator array, due to light spreading at the base of scintillating bars \citep{McConnell2016}.

In the new design an array of completely isolated scintillator elements, each with its own PMT readout, was employed. The number of array elements was reduced to 49. Moreover, the Compton imaging capability was added by including a depth measurement, within each scintillator element, to~allow~GRBs localization. This capability allows also to reduce the cosmic diffuse $\gamma$-ray background.

\subsubsection{SPHiNX: Segmented Polarimeter for High eNergy X-Rays}
SPHiNX \cite{Xie2018} is a proposal to the Swedish Space Agency that underwent a Phase-A/B1 assessment study.
It comprises 42 triangular units of low $Z$ plastics scintillators grouped in hexagons and readout by PMTs. Around each hexagon there are 120 units of high Z rectangular GAGG (Gd$_3$Al$_2$Ga$_3$O$_{12}$) scintillators, which are readout by multi-pixel photon counters (MPPC) to allow the detection of scattered photons. With a field of view of $\pm 60^\circ$ and an energy range between 50--500~keV, it is optimized to study the GRB prompt emission.

\subsubsection{Scattering Polarimetry CdTe/CZT}

New different configurations based on CdTe/CZT detectors \cite{Moita2018} for high-energy polarimetry are under exploration. They include 2D and 3D CZT/CdTe spectroscopic imagers with coincidence readout logic to handle scattering events and to perform simultaneously polarization, spectroscopy, imaging, and timing measurements. Particularly interesting is the development of Laue lenses that would allow a high energy wide band-pass.
In CdTe and CZT detectors the mass attenuation coefficcient\footnote{Derived from \url{https://physics.nist.gov/PhysRefData/Xcom/html/xcom1.html}.} of Compton scattering equals the photoelectric absorption at about 250 keV (see Figure~\ref{fig:scattcrossection} as a comparison with plastic scintillators for which the energy threshold is about 20~keV). Therefore, at lower energy the scattering process is less efficient with respect to photoelectric absorption.

Recently, the AstroSat CZT imager measured the Crab pulsar and nebula polarization in the 100--380~keV energy band \citep{Vadawale2018a}. The AstroSat CZT imager is a coded aperture telescope designed for hard X-ray observations, calibrated also on ground for polarization measurements \citep{Vadawale2015}.
It consists of a pixilated detector plane with a geometric area of 976 cm$^2$ with a pixel thickness of 5 mm and a size of 2.5~mm$\times$2.5~mm. The polarization measurement is performed by detecting coincident events
among neighbour pixels. In the 100-380 keV energy range the recoil electron has an energy much lower than the scattered photon. Therefore, it is assumed that the pixel with the lower energy deposition is the scattering pixel, while the second pixel is the absorbing one. Pixels are squared and the instrument does not rotate.
The polarization dependence of the off-pulse polarization claimed by AstroSat was not confirmed by PoGo+ in a partially overlapping energy band \citep{Chauvin2018}.

\section{Conclusions}

In  the last years X-ray polarimetry in astronomy started to be a crowded field of new theoretical studies, instrument designs and also detectors, which flew on board balloon experiments or demonstrative missions.
Many different physical processes are responsible for the emission of polarized X-rays. These processes involve particle acceleration, emission in strong and ultra-strong magnetic fields, scattering on aspherical geometries and also fundamental physics. The panorama~of astrophysical sources spans from galactic to extragalactic ones and from point like (BHs, NS) to extended (PWNe, SNRs, Molecular Clouds).
Some of the new polarimeters under assessment or development (PolariS, IXPE, eXTP, updated version of GRAPE) have an imaging capability that would allow for the first time to measure the polarization of different regions of extended sources, thus better constraining the parameters of source models. Among these polarimeters, IXPE and eXTP are based on the GPD technology, which allows a quite fine imaging, at the level of some tens of arcseconds. Their imaging capability is currently limited by the optics performance, not by the detector itself.
The~detector technologies available in the quite large energy range, from $\lesssim$1~keV to hundreds~of~keV,\linebreak are based on diffraction onto multilayer mirrors (or crystals), photoelectric effect and scattering (Thomson or Compton).
New activities are planned for the near future and a \textit{critical mass} (among Europe, USA, Japan, China, India etc.) in the scientific community starts to be relevant to sustain larger new projects.
In this framework, particularly relevant is the approval of the Imaging X-ray Polarimetry Explorer (IXPE) that represents the first step towards these larger activities and will make X-ray polarimetry a valuable observational tool worth to be used as well as imaging, spectroscopy and~timing.

\vspace{6pt}
%
%
\acknowledgments{I would like to express my special thancks to the organisers of the ``Alsatian Workshop on X-ray Polarimetry'' held in Strasburg (Fance) the 13--15 November 2017 for inviting me to discuss the topics~of this paper about future missions for X-ray polarimetry.
I would like also to express my special thancks to the agreement ASI-INAF No. 2017-12-H.0 for the participation to the IXPE (Imaging X-ray Polarimeter Explorer) mission approved by NASA for having funded my participation in the meeting.}

\conflictsofinterest{The author is a member of the IXPE Team.}
\reftitle{References}



\end{document}